\documentclass[12pt]{iopart}

\usepackage{iopams}
\usepackage{graphicx}

\newcommand{\Xp}[2]{\Xi^{+}_#1 \left(#2\right)}
\newcommand{\Xm}[2]{\Xi^{-}_#1 \left(#2\right)}

\begin{document}
\title{Fundamental measure theory for lattice fluids with hard core
interactions}

\author{Luis Lafuente and Jos\'e A Cuesta}
\address{Grupo Interdisciplinar de Sistemas Complicados (GISC),
Departamento de Matem\'aticas, Universidad Carlos III de Madrid,
28911 Legan\'es, Madrid, Spain}

\eads{\mailto{llafuent@math.uc3m.es} and \mailto{cuesta@math.uc3m.es}}

\begin{abstract}
We present the extension of Rosenfeld's fundamental measure theory
to lattice models by constructing a density functional for
$d$--dimensional mixtures of parallel hard hypercubes on a
simple hypercubic lattice. The one--dimensional case is exactly
solvable and two cases must be distinguished: all the species
with the same length parity (additive mixture), and arbitrary
length parity (nonadditive mixture). At the best of our knowledge,
this is the first time that the latter case is considered. Based
on the one--dimensional exact functional form, we propose the extension
to higher dimensions by generalizing the zero--dimensional
cavities method to lattice models. This assures the functional to
have correct dimensional crossovers to any lower dimension, including
the exact zero--dimensional limit. Some applications of the functional
to particular systems are also shown.
\end{abstract}

%
\submitto{\JPCM}
\maketitle

\eqnobysec

\section{Introduction}
\label{sec1}
The spirit that has traditionally guided the construction of
approximate density functionals has been to collect as much
information as possible about the fluid and cook up a functional
out of it under reasonable simple assumptions. This is the
idea behind widely used approximations such as the weighted
density or the effective liquid approximations,
which produce a functional out of the Helmholtz free energy
and direct correlation function of the fluid (see Evans 1992
for a review), or even more
recent approaches (Zhou 2001a, 2001b) which in addition incorporate
the pair correlation function. As an alternative to this there
stands Rosenfeld's fundamental measure (FM) theory, whose aim is
to construct the density functional upon geometrical grounds,
and to obtain as output all the ingredients that classical
approaches need as input. Since Rosenfeld's pioneering work
for hard spheres (Rosenfeld 1989), later extended approximately
to nonspherical objects (Rosenfeld 1994), this theory has
developed far from its roots,
giving rise to functionals for hard spheres that reproduce
extraordinarily well the structure of the crystal phase
(Tarazona 2000), functionals for the fluid mixture
of parallel hard cubes (Cuesta and Mart\'{\i}nez--Rat\'on 1997a,
1997b) which have been applied to study entropic demixing
(Mart\'{\i}nez--Rat\'on and Cuesta 1998, 1999) and even functionals
for soft interactions (Schmidt 1999) as well as for nonadditive
mixtures (Schmidt 2000a, 2000b).

Rosenfeld's original approach (Rosenfeld 1989), applied to hard
spheres, consisted of three steps:
(i) decomposing the Mayer function into a sum of
convolutions of one--particle measures;
(ii) using these measures to define a set of weighted densities 
and assuming that the density
functional depends on the density profile only through them, and
(iii) determining the functional form by resorting to
scaled--particle theory.
The functional so obtained had an interesting
functional structure and worked well when applied to problems such
as the adsorption profiles at hard walls. However it dramatically
failed the test of freezing because of some builtin singularities,
which could be repaired by imposing exact dimensional reduction of
the functional to {\em zero--dimensional} cavities, i.e.\ cavities
which do not hold more than one particle (Rosenfeld \etal 1996,
1997). This idea was later exploited by Tarazona and Rosenfeld
(1997) and Tarazona (2000) as a more fundamental idea to construct
a functional free of divergences. The method amounts to extend
to higher dimensions the exact form of the one--dimensional hard--rod
functional (Percus 1976) and then add the appropriate extra
terms (with a form dictated by the original FM functional structure)
so as to cancel whatever exceeds the exact result when the functional
is reduced to properly tailored zero--dimensional cavities. This
method avoids the use of scaled--particle theory while at the same
time imposes a first--principles constraint (the zero--dimensional
reduction) to the functional. It is then not surprising that
the resulting functional performs much better than the original
one, for instance in describing the crystal phase (Tarazona 2000).

The FM functional for the fluid of parallel hard cubes was first
obtained using Rosenfeld's original method (Cuesta 1996)
supplemented with the 
exact zero--dimensional reduction requirement (Cuesta and
Mart\'{\i}nez--Rat\'on 1997a, 1997b). But this system
seems to be more suitable for
FM theory than the hard--sphere one, by at least two reasons: first, 
because the functional turns out to be derivable through an
extraordinarily simple recipe, and second, because, as we will
show for its lattice counterpart, the method of cavities leads
precisely to that
functional. And this remains true not only for the monocomponent
system but also for a mixture, where the FM functional for
hard spheres shows its more prominent weaknesses (Cuesta \etal 2002).

In principle nothing precludes us from extending this theory to
lattice systems. However, when one naively tries to do this
immediately runs into difficulties. To illustrate them let us
suppose we want to follow Ronsenfeld's method. To begin with,
there is no equivalent to scaled--particle theory for lattice
fluids (Soto--Campos \etal 1999) and it is not clear how to
devise it, so this makes step (iii) of the derivation infeasible.
But there is even more: the decomposition of the Mayer function is
ambiguous in the sense that there is not a unique way to define
the fundamental measures. Take, for instance, the one--dimensional
fluid of hard rods in the continuum. The Fourier transform of its
Mayer function is
\[
-\hat f_{ij}(q)=2\,\frac{\sin(a_i+a_j)q}{q}=
2\cos a_iq\frac{\sin a_jq}{q}+2\cos a_jq\frac{\sin a_iq}{q},
\]
where the indices $i,j$ are introduced to distinguish particles
and $2a_i$ is the rod length of particle $i$. As it is clear from
this expression, the decomposition suggests itself. For the lattice
counterpart of this model we have, however,
\[
-\hat f_{ij}(q)=\frac{\sin(a_i+a_j-\frac{1}{2})q}{\sin(q/2)};
\]
without being too imaginative, there are several ways of decomposing
$\hat f_{ij}(q)$, all equally plausible:
\begin{eqnarray*}
-\hat f_{ij}(q)&= \frac{\sin(a_i-\frac{1}{2})q}{\sin(q/2)}\cos a_jq+
\frac{\sin a_jq}{\sin(q/2)}\cos(a_i-{\textstyle\frac{1}{2}})q \\
&= \frac{\sin(a_i-\frac{1}{2})q}{\sin(q/2)}\cos a_jq+
\frac{\sin(a_j-\frac{1}{2})q}{\sin(q/2)}\cos a_iq \\
&\quad +\cos a_iq\cos a_jq+\sin a_iq\sin a_jq \\
&= \frac{\sin(a_i-\frac{1}{4})q}{\sin(q/2)}
\cos(a_j-{\textstyle\frac{1}{4}})q+
\frac{\sin(a_j-\frac{1}{4})q}{\sin(q/2)}
\cos(a_i-{\textstyle\frac{1}{4}})q,
\end{eqnarray*}
and many other possibilities. Contrary to the continuum case, this
gives no clue as to how to define the weighted densities.

An alternative to this approach 
is to use the method of cavities, but there is also
a problem here. The exact one--dimensional functional (Robledo 1980,
Robledo and Varea 1982, Buschle \etal 2000a)
is very different in its structure from the continuum one, so
there is no equivalent to defining convolutions of delta--shells
with appropriate kernels (we will see later that this has no
counterpart in a lattice system). So here again we do not have
a guide to follow.

The method we have employed to achieve the extension and to provide
the right way to define what could be referred to as {\em lattice
fundamental measure theory} (LFMT), rests on two pillars: first, an
appropriate analysis of the extension to mixtures of the exact
one--dimensional functional of hard rods (which, to our knowledge,
we present for the first time in this article), 
and second, a similarity, in the
case of cubic lattices, with the fluid of parallel hard cubes.
After devising the right way to define LFMT we will show how to
make sense of the theory of cavities for lattice fluids. A further
byproduct of this theory is that it would allow for an alternative way
to define a scaled--particle theory for lattice fluids were we able to 
fill in step (iii) of Rosenfeld's method in such a way as to get the
same functional. We are currently working along this line.

The structure of the article goes as follows. In \sref{sec2} we
analyze the mixture of hard rods on a lattice. This means considering
two cases: (a) all rod diameters having the same parity, and (b)
some having odd and some even diameter. The former corresponds to
an additive mixture, while the latter is nonadditive. The
functional for case (a) is derived first, following Vanderlick's \etal
(1989) derivation for the continuum; this method cannot be applied
to case (b), which is then obtained through an appropriate mapping
to an additive mixture in a particular external field. The way
in which those exact functionals are finally written suggests, by
analogy with the fluid of parallel hard cubes, the extension to
higher dimensions (only in cubic lattices). This is carried out in
\sref{sec3}, where it is
also proven that this extension has a correct dimensional reduction
to any lower dimension. It is then show how to extend the method
of cavities for these lattice models. Finally, in \sref{sec4}, we present
some results from applying the obtained functional to some
particular two and three--dimensional systems.

\section{Hard--rod lattice fluid revisited}
\label{sec2}
The exact density functional for the continuum fluid of hard rods
was first obtained by Percus (1976) and extended to multicomponent
mixtures by Vanderlick \etal (1989). Robledo (1980) and Robledo and
Varea (1981) derived the exact functional for the lattice fluid of
hard rods by means of potential--distribution theory (Widom 1978),
but they focused on the continuum limit. Buschle \etal (2000b) have
recently dealt with the general problem of the derivation of exact density
functionals for one--dimensional lattice gases with finite--range pairwise
interaction. They made use of a generalized Markov property satisfied
by the conditional particle distribution probabilities. In this way,
they rederived (Buschle \etal 2000a) the functional for the monocomponent
hard--rod lattice fluid.

In this section, we are going to derive the exact density functional
of a multicomponent lattice fluid of hard rods. The starting point
will be the adaptation to lattice models of 
Vanderlick's \etal (1989) derivation for the continuum.

Let us consider a system of hard rods with different sizes, such that
the interaction between $\alpha$--type and $\alpha'$--type rods at
positions $s$ and $s'$, respectively, on a one--dimensional lattice
(i.e.\ $s,s'\in \mathbb{Z}$) is given by
\begin{equation}
\label{int}
\phi_{\alpha \alpha'}(s,s')=\cases{0&if $|s-s'| \ge 
\sigma_{\alpha \alpha'}$,\\
\infty&if $|s-s'| < \sigma_{\alpha \alpha'}$,\\}
\end{equation}
where $\sigma_{\alpha \alpha'}=\frac{1}{2}(\sigma_{\alpha}+
\sigma_{\alpha'})$, $\sigma_{\alpha}\in \mathbb{N}$ being the diameter
of an $\alpha$--type rod. Because of the particle ordering in
one--dimensional systems, the pairwise interaction will appear in the
grand canonical partition function $\Xi$ via the modified Boltzmann
factor, defined by
\begin{equation}
\label{e}
e_{\alpha \alpha'}(s,s') \equiv \rme^{-\beta
\phi_{\alpha \alpha'}(s,s')} \Theta(s-s')=\Theta(s-s'-\sigma_{\alpha
\alpha'}),
\end{equation}
where $\Theta(s)=1$ if $s \geq 0$ and $0$ otherwise, and $\beta=1/kT$
with $T$ the temperature and $k$ the Boltzmann constant. Note that
this eliminates the combinatorial factor $1/N!$ of the partition
function. Let us also suppose that every species $\alpha$ is in chemical
equilibrium with a particle reservoir of fixed chemical potential
$\mu_{\alpha}$ and that over an $\alpha$--type particle at position
$s\in \mathbb{Z}$ is acting the external potential $u_{\alpha}(s)$.
In the same way, the external and chemical potentials will only appear
in $\Xi$ via
\begin{equation}
w_{\alpha \alpha'}(s,s') \equiv \rme^{\beta\left[
\mu_{\alpha}-u_{\alpha}(s)\right]} \delta_{\alpha \alpha'}
\delta_{s s'} \equiv w_{\alpha}(s) \delta_{\alpha \alpha'} \delta_{s s'},
\end{equation}
where $\delta_{\alpha \alpha'}$ is the Kronecker delta.
With the above definitions we can write the grand canonical partition
function as
\begin{equation}
\label{gpf1}
\Xi=1+\sum_{N=1}^{\infty} \Tr \left[
w_{\alpha_1}(s_1) \prod_{i=1}^{N-1} e_{\alpha_i \alpha_{i+1}}(s_i,s_{i+1})
w_{\alpha_{i+1}}(s_{i+1}) \right],
\end{equation}
where $\Tr \equiv \sum_{\{(\alpha_i,s_i)\}}$ sums over all
indices $(\alpha,s)$ appearing in the expression, $\alpha_i$
denotes the species of the $i$th particle and $s_i \in \mathbb{Z}$ its
position. Considering $e$ and $w$ as operators defined over the
$(\alpha,s)$--space, the term under the sum over $N$ is just
$\Tr \left\{\left[w(ew)^{N-1}\right]_{\alpha \alpha'}(s,s')\right\}$; 
thus we can perform the sum and rewrite \eref{gpf1} as
\begin{equation}
\label{gpf2}
\Xi=1+\Tr \left\{ \left[w(I-ew)^{-1}\right]_{\alpha \alpha'}(s,s') \right\},
\end{equation}
$I$ denoting the identity operator in the $(\alpha,s)$--space.

The density profile of species $\alpha$, $\rho_{\alpha}(s)$,
can be expressed as the functional derivative (Hansen and McDonald 1990)
\begin{equation}
\label{dp1}
\rho_{\alpha}(s)=\frac{w_{\alpha}(s)}{\Xi}\frac{\partial \Xi}
{\partial w_{\alpha}(s)},
\end{equation}
which in this case, given the discrete nature of $s$, is a partial
derivative. Using \eref{gpf2} for $\Xi$ in \eref{dp1}, we can write
$\rho_{\alpha}(s)$ as
\begin{equation}
\label{dp2}
\rho_{\alpha}(s)=\frac{\Xm{\alpha}{s} w_{\alpha}(s)
\Xp{\alpha}{s}}{\Xi},
\end{equation}
where we have introduced the {\it truncated partition functions} defined as
\begin{eqnarray}
\label{t1gpf}
\Xm{\alpha}{s} \equiv \sum_{(\alpha',s')}
(I-ew)^{-1}_{\alpha \alpha'}(s,s'),\\
\label{t2gpf}
\Xp{\alpha}{s} \equiv \sum_{(\alpha',s')}
(I-we)^{-1}_{\alpha' \alpha}(s',s).
\end{eqnarray}
{}From the above definitions and multiplying by $(I-ew)$ to the left
of $\Xm{\alpha}{s}$ and by $(I-we)$ to the right of $\Xp{\alpha}{s}$,
it is straightforward to obtain the following recursive relations
satisfied by the truncated partition functions
\begin{equation}
\label{rec1}
\eqalign{
\Xm{\alpha}{s}=1+\sum_{(\alpha',s')} (ew)_{\alpha \alpha'}(s,s')
\Xm{{\alpha'}}{s'},\\
\Xp{\alpha}{s}=1+\sum_{(\alpha',s')} \Xp{{\alpha'}}{s'}
(we)_{\alpha' \alpha}(s',s).
}
\end{equation}
At this point, our original problem may be posed as to solve \eref{dp2}
and \eref{rec1} in such a way that $w_{\alpha}(s)$ could be
expressed as a functional only depending upon the densities
$\rho_{\alpha}(s)$.

As it was noted by Vanderlick \etal (1989) and Percus (1997), the
solvability of \eref{dp2} and \eref{rec1} depends on
the rank of the Boltzmann factor \eref{e} regarded as an operator in the
$(\alpha,s)$--space. For the additive continuum system it can be shown
that this is of rank one, and then it is possible to solve those equations.
On the contrary, for the lattice system we are dealing with, this is not always
true. In this case, due to the discrete nature of the system, we should
distinguish between a mixture of rods all of them with even(odd) diameters
({\it additive mixture}) and one of rods some of which have even diameter
and some odd ({\it nonadditive mixture}). Indeed, for the former case,
$\sigma_{\alpha \alpha'}$ is always an integer and, consequently,
from \eref{int} it follows that the corresponding mixture
is additive. The solution in this case is just a rephrasing of the
resolution of \eref{dp2} and \eref{rec1} reported for the
continuum system (we will sketch it in \sref{additive}).
For the mixed even--odd case, however, $\sigma_{\alpha \alpha'}$ may
take half--integer values. If $\sigma_{\alpha \alpha'}$ is a half--integer
then $|s-s'| \geq \sigma_{\alpha \alpha'}$ in \eref{int} is also
$|s-s'| \geq \sigma_{\alpha \alpha'} + 1/2$, so effectively the
interaction is nonadditive.
In spite of this, it should be noticed that this nonadditivity is
peculiar for the lattice system and disappears in the continuum
limit. For nonadditive mixtures the Boltzmann factor \eref{e}
is not a rank--one operator; nevertheless, we will show (in
\sref{nonadditive}) how to obtain the exact density functional
for that system as a particular case of the additive mixture.

\subsection{Additive mixture}
\label{additive}
Let us consider a mixture of hard rods such that 
$\sigma_{\alpha}=2a_{\alpha}+\epsilon$ with $a_{\alpha}\in \mathbb{N}$
and $\epsilon=0,1$ depending on whether all the rods have even or odd
diameters, respectively. 
In this case, the Boltzmann factor \eref{e} is a rank--one
operator. If we define the vectors
$e^{+}_{\alpha}(s) \equiv \delta_{s,a_\alpha}$ and
$e^{-}_{\alpha}(s)\equiv \Theta(s-a_\alpha-\epsilon)$ and the convolution
$*$ as $f*g(s)\equiv \sum_{r} f(s-r)g(r)$, \eref{e} can be
written as
\begin{equation}
\label{er1}
e_{\alpha \alpha'}(s,s')=e^{-}_\alpha * e^{+}_{\alpha'}(s-s')
= e^{+}_{\alpha} * e^{-}_{\alpha'}(s-s'). 
\end{equation}
Inserting \eref{er1} in \eref{rec1} we obtain
\begin{equation}
\label{rec2}
\eqalign{
\Xm{\alpha}{s}=1+\sum_{r} e^{-}_{\alpha}(s-r) \left[\sum_{(\alpha',s')}
e^{+}_{\alpha'}(r-s') w_{\alpha'}(s') \Xm{{\alpha'}}{s'}\right],\\
\Xp{\alpha}{s}=1+\sum_{r} \left[ \sum_{(\alpha',s')} \Xp{{\alpha'}}{s'}
w_{\alpha'}(s') e^{+}_{\alpha'}(s'-r) \right] e^{-}_{\alpha}(r-s).
}
\end{equation}
We can now use \eref{dp2} to eliminate $w_{\alpha'}(s')$ from \eref{rec2};
besides, as $e^{-}_{\alpha}(s+a_{\alpha})=\Theta(s-\epsilon)$, hence
independent of $\alpha$, it is handy to define 
\begin{eqnarray}
\label{xipm}
\Xi^{\pm}(s) \equiv \Xi^{\pm}_{\alpha}(s \mp a_{\alpha}),\\
\label{rhopm}
\rho^{\pm}(s) \equiv \sum_{\alpha} \rho_{\alpha}(s \pm a_\alpha),
\end{eqnarray}
which allow us to rewrite equations \eref{rec2} in the more suitable form
\begin{equation}
\eqalign{
\Xi^{-}(s)=1+ \Xi \sum_{r=-\infty}^{s-\epsilon}
\frac{\rho^{-}(r)}{\Xi^{+}(r)},\\
\Xi^{+}(s)=1+ \Xi \sum_{r=s+\epsilon}^{\infty}
\frac{\rho^{+}(r)}{\Xi^{-}(r)}.
}
\end{equation}
{}From the above expressions it is straightforward to obtain the
following system of finite difference equations
\begin{equation}
\label{dif1}
\eqalign{
\Delta \Xi^{-}(s)=\Xi \frac{\rho^{-}(s+1-\epsilon)}{\Xi^{+}
(s+1-\epsilon)},\\
\Delta \Xi^{+}(s)=-\Xi \frac{\rho^{+}(s+\epsilon)}{\Xi^{-}
(s+\epsilon)},
}
\end{equation}
with boundary conditions
\begin{eqnarray}
\label{bc}
\eqalign{
\Xi^{\pm}(\mp \infty)=\Xi, \\
\Xi^{\pm}(\pm \infty)=1,
}
\end{eqnarray}
where we have introduced the difference operator $\Delta$, defined
by $\Delta f(s) \equiv f(s+1)-f(s)$. This system is exactly solvable
and is equivalent to the original recursion relations \eref{rec1}.
In order to solve \eref{dif1} we have to combine the two equations of the
system taking into account the discrete Leibnitz rule $\Delta (fg)(s)=f(s+1)
\Delta g(s)+g(s)\Delta f(s)=g(s+1)\Delta f(s)+f(s)\Delta g(s)$.
This yields
\begin{equation}
\Delta \left(\Xi^{-}\Xi^{+}\right)(s)=\Xi \left[ \rho^{-}
(s+1-\epsilon)-\rho^{+}(s+\epsilon)\right],
\end{equation}
whose solution, taking into account the boundary conditions \eref{bc}, is
\begin{equation}
\label{xx1}
\Xi^{-}(s)\Xi^{+}(s)=\Xi \left\{1+\sum_{r=-\infty}^{s-1}
\left[ \rho^{-}(r+1-\epsilon)-\rho^{+}(r+\epsilon) \right] \right\},
\end{equation}
or using the definitions \eref {rhopm},
\begin{equation}
\label{xx2}
\Xi^{-}(s)\Xi^{+}(s)=\Xi \left( 1 - \sum_{\alpha}
\sum_{r=s-a_{\alpha}+1-\epsilon}^{s+a_{\alpha}-1+\epsilon}
\rho_{\alpha}(r) \right).
\end{equation}
This equation permits us to decouple the system \eref{dif1} and solve
it for $\Xi^{\pm}(s)$. After a little bit of algebra and index
manipulation, the solution has the form
\begin{eqnarray}
\label{sol}
\eqalign{
\frac{\Xi^{-}(s)}{\Xi}=\left[ 1 - n^{(0)}(s-\epsilon) \right]
\prod_{r=s-\epsilon}^{\infty} \frac{1-n^{(1)}(r)}{1-n^{(0)}(r)}, \\
\Xi^{+}(s)=\prod_{r=s}^{\infty} \frac{1-n^{(0)}(r)}
{1-n^{(1)}(r)},
}
\end{eqnarray}
where we have introduced the {\it weighted densities}, $n^{(k)}(s)$ 
($k=0,1$), defined as
\begin{equation}
\label{wdadd}
\fl
n^{(k)}(s)=\sum_{\alpha} \omega_{\alpha}^{(k)}*\rho_{\alpha}(s),
\qquad \omega_{\alpha}^{(k)}(s)=\cases{1&if
$-a_{\alpha}-k-\epsilon<s<a_{\alpha}$,\\
0&otherwise. \\}
\end{equation}

With \eref{sol} we have basically solved our problem,
since from \eref{sol} and \eref{bc} we can write the grand
potential, $\beta \Omega \equiv -\ln \Xi$, as
\begin{equation}
\label{omega}
\beta \Omega = \sum_{s \in \mathbb{Z}} \ln \left( \frac{1-n^{(1)}(s)}
{1-n^{(0)}(s)} \right).
\end{equation}
Finally, the free energy density functional, $\beta \mathcal{F}$, defined
as (Evans 1979)
\begin{equation}
\beta \mathcal{F}[\rho_\alpha]=\beta \Omega[\rho_\alpha] + \sum_{\alpha}
\sum_{s\in \mathbb{Z}} \rho_{\alpha}(s) \ln w_{\alpha}(s),
\end{equation}
takes, from \eref{dp2}, \eref{xipm} and \eref{sol}, the form
\begin{eqnarray}
\fl
\beta \mathcal{F}[\rho_{\alpha}]=\sum_{\alpha} \sum_{s \in \mathbb{Z}}
\rho_{\alpha}(s) \ln \rho_{\alpha}(s) + \beta \Omega[\rho_{\alpha}]
-\sum_{\alpha} \sum_{s \in \mathbb{Z}} \rho_{\alpha}(s) 
\nonumber \\
\times
\left[ \sum_{r=s-a_\alpha-\epsilon}^{s+a_{\alpha}-1} 
\ln\left( 1-n^{(1)}(r) \right)
-\sum_{r=s-a_\alpha-\epsilon+1}^{s+a_{\alpha}-1} 
\ln\left( 1-n^{(0)}(r) \right) \right].
\end{eqnarray}
For reasons that will be made clear later, we prefer to write this
functional in the equivalent, but more convenient form,
\begin{equation}
\label{add}
\beta \mathcal{F}[\rho_\alpha]=\beta \mathcal{F}^{\mathrm{id}}[\rho_\alpha]+
\sum_{s\in \mathbb{Z}} \left[ \Phi_0\left( n^{(1)}(s) \right) -
\Phi_0\left( n^{(0)}(s) \right) \right],
\end{equation}
where $\beta \mathcal{F}^{\mathrm{id}}=\sum_{\alpha} \sum_{s\in \mathbb{Z}}
\rho_\alpha (s) [\ln \rho_\alpha (s) -1]$ refers to the ideal part
of the free energy and $\Phi_0(\eta) \equiv \eta+(1-\eta)\ln(1-\eta)$
is the zero--dimensional excess free energy for a zero--dimensional cavity
with mean occupancy $\eta$ (Rosenfeld \etal 1996).

\subsection{Nonadditive mixture}
\label{nonadditive}
\begin{figure}
\begin{center}
\includegraphics*[width=10cm, angle=0]{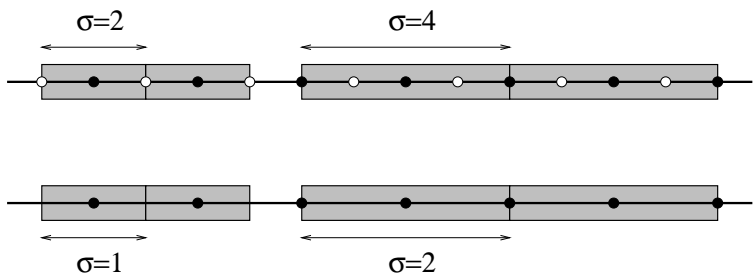}
\end{center}
\caption{\label{fig1} This figure shows how an additive
mixture of hard rods of diameters 2 and 4 can be made to have the same
configurations as 
a nonadditive mixture of rods with diameters 1 and 2.
The positions of the upper system are restricted to be on the black
sites, while the white sites are forbidden.}
\end{figure}

As we have mentioned above, the exact density functional for this special
system can be derived as a particular case of the additive functional
\eref{add}. Let us suppose that we want to study a system of hard rods
whose diameters are $\sigma_{\alpha}=2 a_{\alpha}+\epsilon_{\alpha}$,
where $a_{\alpha} \in \mathbb{N}$ and $\epsilon_{\alpha}=0,1$. Because of
the $\alpha$ dependence in $\epsilon_{\alpha}$, the factorization
\eref{er1} cannot be applied to \eref{e}. This difficulty can be
overcome if we notice (see \fref{fig1}) that this nonadditive mixture
can be mapped to an additive mixture of hard rods whose diameters are
$\tilde{\sigma}_{\alpha} = 2 \sigma_{\alpha}$, while their positions
are restricted to lay on the even sites of the lattice
(i.e.\ $\tilde{\rho}_{\alpha} (s)=0$ if $s=2r+1$ with
$r\in \mathbb{Z}$; this amounts to set an infinite external potential
on the odd sites, while keeping it arbitrary on the even sites).
Therefore, the density functional for the nonadditive system will arise
by introducing in \eref{add} the density profile
\begin{equation}
\label{adddp}
\tilde{\rho}_{\alpha}(s)=\sum_{r \in \mathbb{Z}} \rho_{\alpha}(r)
\delta_{s,2r},
\end{equation}
where $\rho_{\alpha}(s)$ is the density profile corresponding to the
original nonadditive system. This yields for the weighted densities
\eref{wdadd},
\begin{eqnarray}
\label{wdnon}
\eqalign{
n^{(k)}_0(s)\equiv \tilde{n}^{(k)}(2s)=n^{(k)}_{\mathrm{e}}(s)+
n^{(1)}_{\mathrm{o}}(s-1),\\
n^{(k)}_1(s)\equiv \tilde{n}^{(k)}(2s+1)=n^{(1)}_{\mathrm{e}}(s)+
n^{(k)}_{\mathrm{o}}(s),
}
\end{eqnarray}
where $s\in \mathbb{Z}$, $k=0,1$ and the subindex
$\{{\mathrm{e},\mathrm{o}}\}$ states that the sum over species in
the definition \eref{wdadd} is restricted only to those with even or
odd diameter, respectively. There are thus four instead of two weighted
densities for this special case. An alternative way of expressing these
weighted densities is $n^{(k)}_j(s)=\sum_\alpha
\omega^{(k)}_{j,\alpha}*\rho_\alpha(s)$ $(k,j=0,1)$, if the following
nonadditive weights are introduced
\begin{equation}
\label{wnon}
\omega^{(k)}_{j,\alpha}(s)\equiv \cases{\omega^{(j+k(1-j))}_\alpha(s)&
if $\sigma_\alpha$ is even,\\
\omega^{(1-j+kj)}_\alpha(s-1+j)&if $\sigma_\alpha$ is odd.\\}
\end{equation}
With these definitions, the excess
free energy density functional for the nonadditive mixture (the ideal
part remains the same) adopts the form
\begin{eqnarray}
\label{nonad}
\fl
\beta \mathcal{F}^{\mathrm{ex}}[\rho_\alpha]=\sum_{s\in \mathbb{Z}}\left[
\Phi_0\left(n^{(1)}_1(s)\right) + \Phi_0\left(n^{(1)}_0(s)\right)
-\Phi_0\left(n^{(0)}_1(s)\right) - \Phi_0\left(n^{(0)}_0(s)
\right) \right].
\end{eqnarray}

As far as we know, functionals \eref{add} and \eref{nonad} have never been
reported in the literature. The additive case is just an exercise of
rewriting Vanderlick's \etal (1989) original derivation for the continuum,
but this nonadditive case cannot be directly obtained with their method.
Besides, even the functional \eref{add} (which for the monocomponent
case reduces to that obtained several times in the literature) has been
written in a form that will make of the obtention of an approximate
functional in higher dimensions an easy task.

\section{From 0D cavities to higher dimensions}
\label{sec3}
Cuesta and Mart\'{\i}nez--Rat\'on (1997a) proved that the one--dimensional
excess free energy functional in the continuum model can be generated 
from the zero--dimensional one by just applying
a differential operator with respect to the particle sizes. The
form in which the functionals \eref{add} and \eref{nonad} have been
written proves that this is also true for the discrete case, but
now, due to the discrete nature of the system, the differential operator
becomes a difference operator. This suggests rewriting \eref{add} and
\eref{nonad} as
\begin{equation}
\label{add2}
\fl
\beta \mathcal{F}^{\mathrm{ex}}[\rho_\alpha]=\sum_{s\in \mathbb{Z}}
\mathcal{D}_k \Phi_0\left(n^{(k)}(s)\right), \qquad
\beta \mathcal{F}^{\mathrm{ex}}[\rho_\alpha]=\sum_{s\in \mathbb{Z}}
\mathcal{D}_k \left[\sum_{j=0,1} \Phi_0\left(n^{(k)}_j(s)\right)\right],
\end{equation} 
respectively. Here $\mathcal{D}_k$ is the difference operator
$\mathcal{D}_k f(k)\equiv f(1)-f(0)$. Cuesta and Mart\'{\i}nez--Rat\'on 
(1997a) also proved
that the FM functional for the
$d$--dimensional system of parallel hard cubes followed the same
simple recipe, i.e.\ it can be generated from the zero--dimensional
functional by successive applications of a differential operator
with respect to the particle size along every coordinate axis.
The functional so obtained satisfies the zero--dimensional reduction
requirement. For the equivalent lattice model it seems natural to 
propose a similar recipe. Then, taking into account the continuum 
analogue and \eref{add2}, our proposal for the FM
functional of the $d$--dimensional discrete parallel hard cube model is
\begin{equation}
\label{add3}
\beta \mathcal{F}^{\mathrm{ex}}[\rho_\alpha] = \sum_{\mathbf{s}\in
\mathbb{Z}^{d}} \mathcal{D}_{\mathbf{k}} \Phi_0\left(n^{(\mathbf{k})}
(\mathbf{s})\right),
\end{equation}
for the additive case, and
\begin{equation}
\label{nonad3}
\beta \mathcal{F}^{\mathrm{ex}}[\rho_\alpha] = \sum_{\mathbf{s}\in
\mathbb{Z}^{d}} \mathcal{D}_{\mathbf{k}} \left[ \sum_{\mathbf{j}\in
\{0,1\}^d} \Phi_0\left(n^{(\mathbf{k})}_{\mathbf{j}}(\mathbf{s})\right)
\right],
\end{equation}
for the nonadditive one,
where $\mathbf{k}\equiv (k_1,\ldots,k_d)$, $\mathbf{j}\equiv
(j_1,\ldots,j_d)$ are vector indices, $\mathcal{D}_{\mathbf{k}}
\equiv \prod_{i=1}^d \mathcal{D}_{k_i}$ is the difference operator 
and the weighted densities are
defined as $n^{(\mathbf{k})}(\mathbf{s}) \equiv
\sum_{\alpha} \omega^{(\mathbf{k})}_\alpha * \rho_\alpha(\mathbf{s})$
with $\omega^{(\mathbf{k})}_\alpha(\mathbf{s}) \equiv \prod_{i=1}^d
\omega^{(k_i)}_\alpha(s_i)$, for the additive system, and as
$n^{(\mathbf{k})}_{\mathbf{j}}(\mathbf{s}) \equiv \sum_{\alpha}
\omega^{(\mathbf{k})}_{\mathbf{j},\alpha} * \rho_\alpha(\mathbf{s})$
with $\omega^{(\mathbf{k})}_{\mathbf{j},\alpha}(\mathbf{s}) \equiv
\prod_{i=1}^d \omega^{(k_i)}_{j_i,\alpha}(s_i)$, for the nonadditive
one. 

Functionals
\eref{add3} and \eref{nonad3} are the main results of this work since
they constitute, to the best of our knowledge, the first
FM functionals for a lattice model. In this sense,
it can be checked that (i) they depend on the density profile through a
set of weighted densities defined from one--particle measures which
are compatible with the decomposition of the Mayer function into a sum
of their convolutions; (ii) the virial expansion of the direct correlation
function is exact up to first order in the density, and (iii) they 
consistently reduce to any lower dimension down to $d=0$. Also, it must 
be noticed that their continuum counterpart is recovered in the continuum 
limit of vanishing lattice spacing $\delta\to 0$ and infinite size
$\sigma_{\alpha}\to\infty$ with $\sigma_{\alpha}\delta=\mbox{const.}$ 
(notice that the additive--nonadditive differentiation becomes
immaterial in this limit). Points (i) and (ii) are easy to check, so
we will focus in what follows in the crucial point (iii) and its
connection with an alternative way of obtaining functionals
\eref{add3} and \eref{nonad3}.

As it was mentioned in the introduction, there is an 
alternative method to Rosenfeld's original one to construct a 
FM functional. This method will be
referred to as the {\em zero--dimensional cavities method}
(Rosenfeld \etal 1996, Tarazona and Rosenfeld 1997, Tarazona 2000)
because it is based on the exact dimensional reduction of the 
functional to $d=0$.
Pursuing the same idea but having in mind the form of functionals
\eref{add} and \eref{nonad}, we have devised a similar method
suitable for lattice models. The method amounts to (a) directly
extending the exact zero--dimensional functional to higher dimensions
(note here a difference with the continuum case (Tarazona and Rosenfeld 
1997), where it is the one--dimensional functional what is extended 
to higher
dimensions), and then (b) adding the appropriate extra terms (in the 
way suggested by \eref{add} and \eref{nonad}) in order for the 
functional to recover
the exact form when applied to zero--dimensional cavities.

\begin{figure}
\begin{center}
\includegraphics*[width=12cm, angle=0]{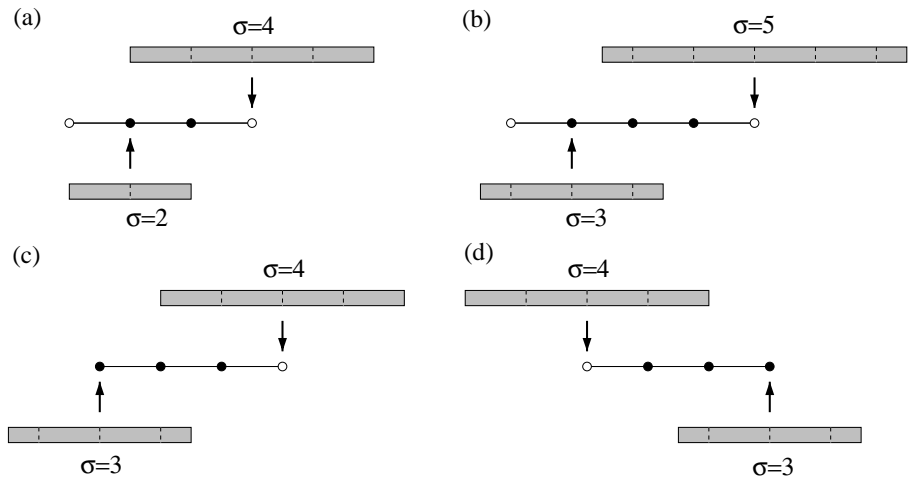}
\end{center}
\caption{\label{fig2} Maximal cavities for binary mixtures in
$d=1$. Large particles can be placed at any point, black or white,
but small ones can only
be placed at black sites. The figure shows the additive even
(a) and odd (b) cases, as well as the two equivalent cavities
for one nonadditive case (c) and (d).}
\end{figure} 

To illustrate this procedure we must first recall that we
are dealing with a mixture, sometimes nonadditive, and, contrary
to the monocomponent case, it is not very clear what a 
zero--dimensional cavity means for a mixture. So let us start
by defining what we will understand by that. In the monocomponent 
case, a zero--dimensional cavity is simply a set of connected
points in the $d$--dimensional lattice such that if a particle
(its center of mass) is placed at one of them no other particle 
can be placed at any other point of this set. For 
one--dimensional hard rods of length $2a$,
the largest zero-dimensional cavity (`maximal' cavity) is the 
interval $[-a+1,a]$, whereas for hard
rods of length $2a+1$ it is $[-a+1,a+1]$. In $d$ dimensions the maximal
cavity is the set $[-a+1,a]^d$ or $[-a+1,a+1]^d$, depending on the
length parity. 

For a mixture, a zero--dimensional cavity can be defined as a 
collection of sets, one for each species, such that if a particle 
of any species occupies one of the points of its corresponding
set, no other particle of the same or different species can be
placed at any point of its corresponding set (\fref{fig2} will
help to clarify this concept). In $d=1$ and in 
the additive case, this means the collection of sets 
$\{[-a_{\alpha}+1,a_{\alpha}]\}_{\alpha}$, if $2a_{\alpha}$ is
the rod length of species $\alpha$ (\fref{fig2}a), or $\{[-a_{\alpha}+1,
a_{\alpha}+1]\}_{\alpha}$, if this rod length is $2a_{\alpha}+1$
(\fref{fig2}b); in the nonadditive case, however, indexing by 
$\alpha$ the species of even length and by $\alpha'$ those of 
odd length, there are two maximal cavities: $\{[-a_{\alpha}+1,
a_{\alpha}],[-a_{\alpha'},a_{\alpha'}]\}_{\alpha,\alpha'}$ 
(\fref{fig2}c) and $\{[-a_{\alpha}+1,a_{\alpha}],[-a_{\alpha'}+1,
a_{\alpha'}+1]\}_{\alpha,\alpha'}$ (\fref{fig2}d).
In $d$ dimensions the cavities for the additive case generalize 
as in the monocomponent case, while in the nonadditive case
there will appear $2^d$ different cavities, depending on which of
the two choices above we select for each dimension.
Because of this proliferation of maximal cavities in the nonadditive
case, it is convenient to discuss both types of mixtures separately.

\subsection{Additive mixture}

For density profiles constrained to be zero outside the points of
a maximal zero-dimensional cavity it is easy to check (by applying
the definition) that the weighted densities \eref{wdadd} satisfy
the relationships
\begin{eqnarray}
\label{rel1d}
\eqalign{
n^{(1)}(s)=n^{(0)}(s+1)\quad&(s < 0),\\
n^{(1)}(s)=n^{(0)}(s)&(s > 0).
}
\end{eqnarray}
These are the keystone of the method. As explained, the method
starts off from the straightforward extension of the 
zero--dimensional excess functional to $d=1$, $\Phi_0^{(1d)}$.
As $n^{(1)}(0)=\eta$,
the occupancy probability of the cavity,
\begin{equation}
\Phi_0^{(1d)}[\rho]=\sum_{s\in\mathbb{Z}}\Phi_0\left(n^{(1)}(s)\right).
\label{exten1d}
\end{equation}
Now we apply this functional to a zero--dimensional cavity and, 
by means of \eref{rel1d} rewrite it as
\begin{eqnarray}
\label{0d1d}
\Phi_0^{(1d)}[\rho]
&=\Phi_0\left(n^{(1)}(0)\right)+\sum_{s<0} \Phi_0\left(n^{(0)}(s+1)\right)
+\sum_{s>0} \Phi_0\left(n^{(0)}(s)\right)\nonumber\\
&=\Phi_0\left(n^{(1)}(0)\right)+\sum_{s\in\mathbb{Z}} \Phi_0\left(
n^{(0)}(s)\right).
\end{eqnarray}
But $\Phi_0\left(n^{(1)}(0)\right)$ at the
r.h.s.\ of \eref{0d1d} is the exact zero--dimensional excess
free energy for the cavity, so the method dictates that in order
to obtain the excess functional for the one--dimensional system we
must substract to the form we started from all the spurious terms.
This leads to
\begin{equation}
\beta \mathcal{F}^{\mathrm{ex}}[\rho]=\sum_{s\in\mathbb{Z}}
\Phi_0\left(n^{(1)}(s)\right)-\sum_{s\in\mathbb{Z}}
\Phi_0\left(n^{(0)}(s)\right),
\end{equation}
which coincides with \eref{add}.

Let us now illustrate how the method works for higher dimensions
as well. For the sake of simplicity we will restrict ourselves to
do it only for $d=2$. The corresponding relationships between the 
weighted densities of density profiles vanishing outside a
zero--dimensional cavity are
\begin{eqnarray}
\label{rel2d}
\eqalign{
n^{(k,1)}(s_1,s_2)=n^{(k,0)}(s_1,s_2+1)\quad&(s_2 < 0), \\
n^{(k,1)}(s_1,s_2)=n^{(k,0)}(s_1,s_2)&(s_2 > 0),
}
\end{eqnarray}
with $k=0,1$, and similar ones if the coordinate axes are 
interchanged. Using the same notation as for the one--dimensional
case and repeating the steps in \eref{0d1d} for $s_2$ we get
\begin{equation}
\fl
\Phi_0^{(2d)}[\rho]=\sum_{\mathbf{s}\in\mathbb{Z}^2}
\Phi_0\left(n^{(1,1)}(\mathbf{s})\right) 
=\sum_{s_1\in\mathbb{Z}} \Phi_0\left(n^{(1,1)}(s_1,0)\right)+
\sum_{\mathbf{s}\in\mathbb{Z}^2} \Phi_0\left(n^{(1,0)}(\mathbf{s})\right).
\label{2d0da}
\end{equation}
Repeating the trick with $\sum_{\mathbf{s}\in\mathbb{Z}^2}
\Phi_0\left(n^{(0,1)}(\mathbf{s})\right)$ we get
\begin{equation}
\label{zero}
\fl
0=\sum_{\mathbf{s}\in\mathbb{Z}^2} \Phi_0\left(n^{(0,1)}(\mathbf{s})\right)
-\sum_{s_1\in\mathbb{Z}} \Phi_0\left(n^{(0,1)}(s_1,0)\right)
-\sum_{\mathbf{s}\in\mathbb{Z}^2} \Phi_0\left(n^{(0,0)}(\mathbf{s})\right).
\end{equation}
Adding \eref{zero} to \eref{2d0da} and noticing that
\begin{equation}
\fl
\sum_{s_1\in\mathbb{Z}} \Phi_0\left(n^{(1,1)}(s_1,0)\right)
-\sum_{s_1\in\mathbb{Z}} \Phi_0\left(n^{(0,1)}(s_1,0)\right)=
\Phi_0\left(n^{(1,1)}(\mathbf{0})\right)
\end{equation}
yields
\begin{eqnarray}
\fl
\Phi_0^{(2d)}[\rho]=\Phi_0\left(n^{(1,1)}(\mathbf{0})\right)
\nonumber \\
+\sum_{\mathbf{s}\in\mathbb{Z}^2}\left[
\Phi_0\left(n^{(1,0)}(\mathbf{s})\right)+
\Phi_0\left(n^{(0,1)}(\mathbf{s})\right)-
\Phi_0\left(n^{(0,0)}(\mathbf{s})\right)\right].
\end{eqnarray}
Again, $\Phi_0\left(n^{(1,1)}(\mathbf{0})\right)$ is the exact 
zero--dimensional excess functional, therefore
\begin{eqnarray}
\fl
\beta \mathcal{F}^{\mathrm{ex}}[\rho]=\sum_{\mathbf{s}\in\mathbb{Z}^2}
\left[\Phi_0\left(n^{(1,1)}(\mathbf{s})\right) \right.
 \nonumber \\
\left.  -\Phi_0\left(n^{(1,0)}(\mathbf{s})\right)
-\Phi_0\left(n^{(0,1)}(\mathbf{s})\right)+
\Phi_0\left(n^{(0,0)}(\mathbf{s})\right)\right],
\end{eqnarray}
which coincides with \eref{add3} for $d=2$. This provides the 
rationale for \eref{add3} for arbitrary dimension, and proves in
passing that functional \eref{add3} has a consistent dimensional
reduction to any lower dimension.

\subsection{Nonadditive mixture}

\begin{figure}
\begin{center}
\includegraphics*[width=10cm, angle=0]{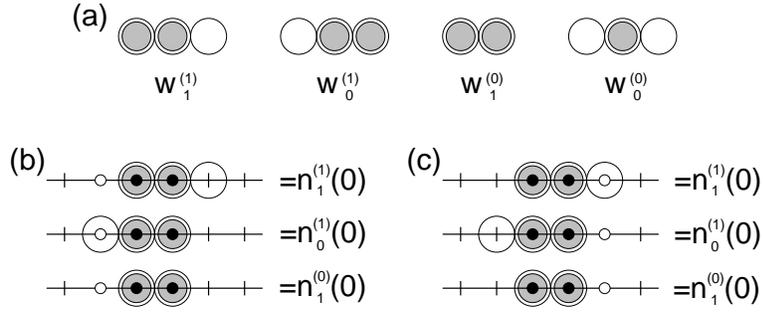}
\end{center}
\caption{\label{fig3} (a) Diagrammatic representation of the weights
corresponding to a binary mixture of rods of lengths $\sigma_1=3$
and $\sigma_2=2$. Circles window points whose density contribute
to the weighted density: empty circles for the large rods and
shaded circles for the small rods. (b) and (c) Result of convoluting 
the weights with the density profiles of zero--dimensional
maximal cavities.}
\end{figure} 

In the additive case it is always the weighted density $n^{(1)}(0)$
what gives the occupancy probability, $\eta$, of the (unique) maximal
zero--dimensional cavity, so extending the zero--dimensional excess
functional is as straightforward as equation \eref{exten1d} shows.
In the nonadditive case, however, this is not as simple. Due to
the degeneracy of maximal cavities, which weighted density yields
the occupancy probability depends on which cavity we choose. So
before proceeding with the construction of the functional we need
a unique expression in terms of the one--dimensional weighted 
densities that provides the excess free energy of {\em any} maximal 
cavity. One such expression is
\begin{equation}
\label{0dnonad}
\Phi_0^{(0d)}[\rho]=\Phi_0\left(n_1^{(1)}(0)\right)+
\Phi_0\left(n_0^{(1)}(0)\right)-
\Phi_0\left(n_1^{(0)}(0)\right),
\label{phi0dnonad}
\end{equation}
In order to make it easy to understand why this is so and the
discussions to come, we have introduced a diagrammatic notation
for a particular case in \fref{fig3}. We will consider a binary
mixture of rods of lengths $\sigma_1=3$ and $\sigma_2=2$, which is
prototypical of this case. For this mixture we represent the
weights as a chain of empty and shaded circles (see \fref{fig3}a). 
The effect of convoluting
a weight with a density profile is to select the density at
those sites overlapped by the empty circles for the large rods
and by the shaded circles for the small rods, and add them up.
The result of this operation is the corresponding weighted density.
The application of these weights to each maximal cavity is depicted
in \fref{fig3}b and c, respectively. \Fref{fig3}b illustrates
that $n_0^{(1)}(0)=\eta$ and $n_1^{(1)}(0)=n_1^{(0)}(0)$, so
\eref{phi0dnonad} reduces to $\Phi_0(\eta)$, as expected. On the
other hand, \fref{fig3}c illustrates that $n_1^{(1)}(0)=\eta$ and 
$n_0^{(1)}(0)=n_1^{(0)}(0)$, so again \eref{phi0dnonad} reduces 
to $\Phi_0(\eta)$.

We can now carry on with the method as in the additive case
and propose
\begin{equation}
\label{1dnonad}
\Phi_0^{(1d)}[\rho]=\sum_{s\in\mathbb{Z}}\left[
\Phi_0\left(n_1^{(1)}(s)\right)+\Phi_0\left(n_0^{(1)}(s)\right)-
\Phi_0\left(n_1^{(0)}(s)\right)\right].
\end{equation}

Whichever the maximal cavity we choose, from definitions \eref{wdnon}
and the relationships \eref{rel1d} the following analogous relationships
between the nonadditive weighted densities hold
\begin{eqnarray}
\label{relnonad}
\eqalign{
n^{(1)}_0(s)=n^{(0)}_1(s),\quad n^{(1)}_1(s)=n^{(0)}_0(s+1)\quad&(s<0),\\
n^{(1)}_0(s)=n^{(0)}_0(s),\quad n^{(1)}_1(s)=n^{(0)}_1(s)&(s>0).
}
\end{eqnarray}
Proceeding as we did for the additive case, we use the above relationships
to rewrite \eref{1dnonad} as
\begin{eqnarray}
\label{0d1dnonad}
\Phi_0^{(1d)}[\rho]
&=\Phi_0^{(0d)}[\rho]+\sum_{s<0} \Phi_0\left(n^{(0)}_0(s+1)\right)
+\sum_{s>0} \Phi_0\left(n^{(0)}_0(s)\right)\nonumber\\
&=\Phi_0^{(0d)}[\rho]+\sum_{s\in\mathbb{Z}} \Phi_0\left(
n^{(0)}_0(s)\right),
\end{eqnarray}
where we have made use of \eref{0dnonad}, which gives the zero--dimensional
excess functional; so the method gives
\begin{eqnarray}
\fl
\beta \mathcal{F}^{\mathrm{ex}}[\rho_\alpha]=\sum_{s\in \mathbb{Z}}\left[
\Phi_0\left(n^{(1)}_1(s)\right) + \Phi_0\left(n^{(1)}_0(s)\right)
-\Phi_0\left(n^{(0)}_1(s)\right) - \Phi_0\left(n^{(0)}_0(s)
\right) \right],
\end{eqnarray}
which coincides with \eref{nonad}.

\begin{figure}
\begin{center}
\includegraphics*[width=11cm, angle=0]{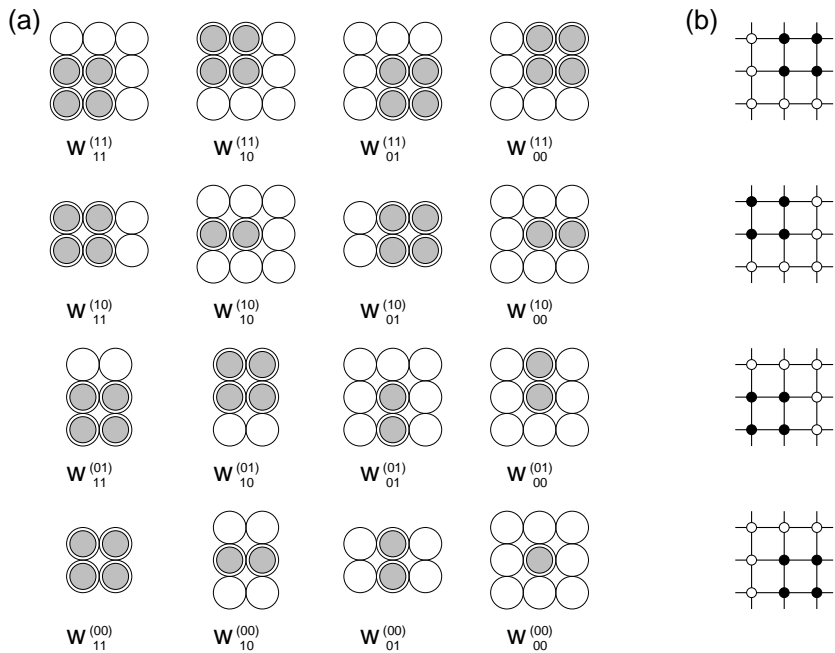}
\end{center}
\caption{\label{fig4} (a) Diagrammatic representation of the weights
corresponding to a binary mixture of squares of edge lengths
$\sigma_1=3$ and $\sigma_2=2$ (see caption of \fref{fig3}a to 
understand the circle notation). (b) Maximal cavities for that system.
Note that the result of convoluting the weights in (a) with the
density profiles of (b), $n^{(\mathbf{k})}_{\mathbf{j}}(\mathbf{0})$,
is obtained by superposing the lower leftmost
shadow circle of the weight with the lower leftmost black point of the cavity.}
\end{figure} 

In order to extend the cavities method to dimensions higher than one,
we have to generalize expression \eref{0dnonad}.
Due to the cubic symmetry of our system this can be done just by
applying the same rule on each coordinate axis. For the sake of simplicity,
we can define
the operator $\mathcal{T}_k^j$ as $\mathcal{T}_k^j f(k,j)\equiv
f(1,1)+f(1,0)-f(0,1)$ and its $d$--dimensional version as
$\mathcal{T}_{\mathbf{k}}^{\mathbf{j}}\equiv \prod_{i=1}^{d}
\mathcal{T}_{k_i}^{j_i}$. Then the natural generalization of
\eref{0dnonad} is simply
\begin{equation}
\label{0dnonad2}
\Phi_0^{(0d)}[\rho]=\mathcal{T}_{\mathbf{k}}^{\mathbf{j}}
\Phi_0\left(n_{\mathbf{j}}^{(\mathbf{k})}(\mathbf{0})\right),
\end{equation}
and the extension of the zero--dimensional excess functional to
$d$ dimensions
\begin{equation}
\label{0dd}
\Phi_0^{(d)}[\rho]=\sum_{\mathbf{s}\in\mathbb{Z}^d} \mathcal{T}_
{\mathbf{k}}^{\mathbf{j}} \Phi_0\left(n_{\mathbf{j}}^{(\mathbf{k})}
(\mathbf{s})\right).
\end{equation}
Again, as a representative of a higher dimension, we will apply the cavities
method for $d=2$. In \fref{fig4} it is shown the diagrammatic
representation of the weights for a binary mixture of squares of
edge lengths $\sigma_1=3$ and $\sigma_2=2$, as well as all maximal
cavities for that system. Expression \eref{0dnonad2} for $d=2$
accounts for all cavities in \fref{fig4}b. For example, if we
take the cavity at the top the application of the weights in
\fref{fig4}a gives $n^{(1,1)}_{0,0}(\mathbf{0})=\eta$,
$n^{(1,1)}_{1,1}(\mathbf{0})=n^{(0,0)}_{1,1}(\mathbf{0})$,
$n^{(1,1)}_{1,0}(\mathbf{0})=n^{(0,1)}_{1,0}(\mathbf{0})$,
$n^{(1,1)}_{0,1}(\mathbf{0})=n^{(1,0)}_{0,1}(\mathbf{0})$ and
$n^{(1,0)}_{1,1}(\mathbf{0})=n^{(0,1)}_{1,1}(\mathbf{0})$,
so \eref{0dnonad2} reduces to
$\Phi_0(\eta)$. By symmetry the same happens for any of the cavities.

Proceeding with the method, from \eref{0dd}
we will start with
\begin{eqnarray}
\label{0d2dnonad}
\fl
\Phi_0^{(2d)}[\rho]=\sum_{\mathbf{s}\in{\mathbb{Z}^2}}\left[
\Phi_0\left(n^{(1,1)}_{1,1}(\mathbf{s})\right)+
\Phi_0\left(n^{(1,1)}_{1,0}(\mathbf{s})\right)-
\Phi_0\left(n^{(1,0)}_{1,1}(\mathbf{s})\right) \right. \nonumber \\
+\Phi_0\left(n^{(1,1)}_{0,1}(\mathbf{s})\right)
+\Phi_0\left(n^{(1,1)}_{0,0}(\mathbf{s})\right)
-\Phi_0\left(n^{(1,0)}_{0,1}(\mathbf{s})\right) \nonumber \\
\left. -\Phi_0\left(n^{(0,1)}_{1,1}(\mathbf{s})\right)
-\Phi_0\left(n^{(0,1)}_{1,0}(\mathbf{s})\right)+
\Phi_0\left(n^{(0,0)}_{1,1}(\mathbf{s})\right)
\right].
\end{eqnarray}
Now, the relationships \eref{relnonad} become
\begin{eqnarray}
\label{rel2nonad}
\fl
\eqalign{
n^{(k,1)}_{j,0}(s_1,s_2)=n^{(k,0)}_{j,1}(s_1,s_2),\quad
n^{(k,1)}_{j,1}(s_1,s_2)=n^{(k,0)}_{j,0}(s_1,s_2+1)\quad&(s_2<0),\\
n^{(k,1)}_{j,0}(s_1,s_2)=n^{(k,0)}_{j,0}(s_1,s_2),\quad
n^{(k,1)}_{j,1}(s_1,s_2)=n^{(k,0)}_{j,1}(s_1,s_2)&(s_2>0),
}
\end{eqnarray}
and those obtained by interchanging the coordinate axes.
As before, by making use of these relationships in \eref{0d2dnonad}
we get
\begin{eqnarray}
\label{0d2daux}
\fl
\Phi_0^{(2d)}[\rho]=\sum_{s_1\in\mathbb{Z}} \mathcal{T}_{\mathbf{k}}^
{\mathbf{j}} \Phi_0\left(n^{(\mathbf{k})}_{\mathbf{j}}(s_1,0)\right)
\nonumber \\
+\sum_{\mathbf{s}\in\mathbb{Z}^2} \left[
\Phi_0\left(n^{(1,0)}_{1,0}(\mathbf{s})\right)+
\Phi_0\left(n^{(1,0)}_{0,0}(\mathbf{s})\right)-
\Phi_0\left(n^{(0,0)}_{1,0}(\mathbf{s})\right)
\right],
\end{eqnarray}
where it must be noticed that the second term in the r.h.s.\ of \eref{0d2daux}
is just $\sum_{\mathbf{s}\in \mathbb{Z}^2} \mathcal{T}_k^j
\Phi_0\left(n^{(k,0)}_{j,0}(\mathbf{s})\right)$. By the same argument, we have
\begin{equation}
\label{0d2daux2}
\fl
0=\sum_{\mathbf{s}\in\mathbb{Z}^2} \mathcal{T}_k^j
\Phi_0\left(n^{(0,k)}_{0,j}(\mathbf{s})\right)-
\sum_{s_1\in\mathbb{Z}} \mathcal{T}_k^j
\Phi_0\left(n^{(0,k)}_{0,j}(s_1,0)\right)-
\sum_{\mathbf{s}\in\mathbb{Z}} \Phi_0\left(n^{(0,0)}_{0,0}(\mathbf{s}).
\right)
\end{equation}
Adding \eref{0d2daux2} to \eref{0d2daux} and noticing that because of the
zero--dimensional reduction of the one--dimensional case proved above
\begin{equation}
\sum_{s_1 \in \mathbb{Z}}\left[ \mathcal{T}_{\mathbf{k}}^{\mathbf{j}}
\Phi_0\left(n^{(\mathbf{k})}_{\mathbf{j}}(s_1,0)\right)-
\mathcal{T}_k^j
\Phi_0\left(n^{(0,k)}_{0,j}(s_1,0)\right)\right]=\Phi_0^{(0d)}[\rho],
\end{equation}
we obtain
\begin{eqnarray}
\fl
\Phi_0^{(2d)}[\rho]=\Phi_0^{(0d)}[\rho]+
\sum_{\mathbf{s}\in\mathbb{Z}^2} \left[
\Phi_0\left(n^{(1,0)}_{1,0}(\mathbf{s})\right)+
\Phi_0\left(n^{(1,0)}_{0,0}(\mathbf{s})\right)
\right] \nonumber \\
+\sum_{\mathbf{s}\in\mathbb{Z}^2} \left[
\Phi_0\left(n^{(0,1)}_{0,1}(\mathbf{s})\right)+
\Phi_0\left(n^{(0,1)}_{0,0}(\mathbf{s})\right)
\right] \nonumber \\
-\sum_{\mathbf{s}\in\mathbb{Z}^2} \left[
\Phi_0\left(n^{(0,0)}_{1,0}(\mathbf{s})\right)
+\Phi_0\left(n^{(0,0)}_{0,1}(\mathbf{s})\right)
+\Phi_0\left(n^{(0,0)}_{0,0}(\mathbf{s})\right) \right].
\end{eqnarray}
Thus, the cavities method produces
\begin{eqnarray}
\fl
\beta \mathcal{F}^{\mathrm{ex}}[\rho]=\Phi_0^{(2d)}[\rho]-
\sum_{\mathbf{s}\in\mathbb{Z}^2} \left[
\Phi_0\left(n^{(1,0)}_{1,0}(\mathbf{s})\right)+
\Phi_0\left(n^{(1,0)}_{0,0}(\mathbf{s})\right)
\right] \nonumber \\
-\sum_{\mathbf{s}\in\mathbb{Z}^2} \left[
\Phi_0\left(n^{(0,1)}_{0,1}(\mathbf{s})\right)+
\Phi_0\left(n^{(0,1)}_{0,0}(\mathbf{s})\right)
\right] \nonumber \\
+\sum_{\mathbf{s}\in\mathbb{Z}^2} \left[
\Phi_0\left(n^{(0,0)}_{1,0}(\mathbf{s})\right)
+\Phi_0\left(n^{(0,0)}_{0,1}(\mathbf{s})\right)
+\Phi_0\left(n^{(0,0)}_{0,0}(\mathbf{s})\right) \right],
\end{eqnarray}
which is \eref{nonad3} for $d=2$.

The extension of this argument to arbitrary dimension proves that
\eref{nonad3} is the functional produced by
the cavities method. In addition, it is also proved its consistent
dimensional reduction to any lower dimension.

\section{Some applications}
\label{sec4}
In order to show the applicability
of the theory exposed in the previous sections, in this one
we present the results obtained for some particular
two and three--dimensional systems. 
\subsection{Hard square lattice gas}
The hard square lattice gas is just the lattice gas with first and
second nearest--neighbour exclusion in the simple square lattice.
This system has been widely studied (Bellemans and Nigam 1967,
Ree and Chesnut 1967, Nisbet and Farquhar 1974, Slotte 1983)
but no definitive conclusion has been
reached about its phase behaviour. The results reported cover
the whole spectrum, depending on which theory has been
used to study it. Some authors have claimed it to have
a second order transition (Bellemans and Nigam 1967, Slotte 1982),
others a weaker (third order) transition (Bellemans and Nigam 1967,
Ree and Chesnut 1967) and others even no transition at all (Bellemans
and Nigam 1967, Nisbet and Farquhar 1974, Slotte 1982).
Nowadays, the only thing that seems
clear is the structure of the ordered phase: periodic
along one coordinate axis while
uniform along the other (i.e.\ columnar).
However, no simulations have ever
been performed in order to clarify the disputed order of the transition.

\begin{figure}
\begin{center}
\includegraphics*[width=7cm, angle=-90]{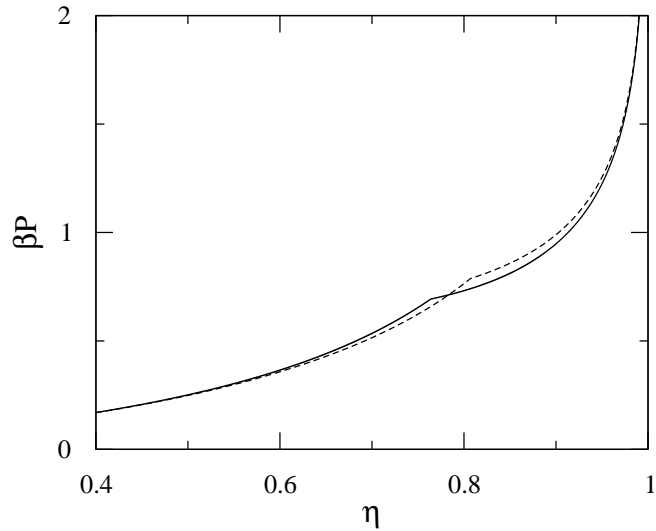}
\end{center}
\caption{\label{fig5} Equation of state (reduced pressure,
$\beta P$, vs. packing fraction, $\eta$) from FM theory
(solid line) and from Rushbrooke and Scoins's method (dashed line).
}
\end{figure}

The result of FM theory is
shown in \fref{fig5}. The system is found to have a second order
transition from a fluid phase to a columnar phase. The value of
the packing fraction at the transition is $\eta_\mathrm{c}=3-\sqrt{5}=
0.764$ and the chemical potential at that point $(\beta \mu)_{\mathrm{c}}=
2.41$. This result is compatible with that obtained by Bellemans and Nigam
(1967) using Rushbrooke and Scoins's (1955) method. They also found
a second order transition, but at $\eta_\mathrm{c}=0.807$ and
$(\beta \mu)_{\mathrm{c}}=2.85$. The equation of state they obtained
with that method is compared with the FM result
in \fref{fig5}. There is good agreement between both
theories, being only in the critical region where the two curves
deviate from each other.

It is worth mentioning that within the FM approximation
this system is analytically solvable.
Omitting the calculations and denoting $L(x)\equiv x \ln x$, we have
obtained for the Helmholtz free energy density of the fluid branch
\begin{equation}
\beta \Phi(\eta)=L(\eta/4)+L(1-\eta/4)-2 L(1-\eta/2)+L(1-\eta),
\end{equation} 
and for the columnar branch
\begin{eqnarray}
\fl
\beta \Phi(\eta)=\frac{1}{2}\left[L(\eta_1/2)+L(1-\eta_1/2)
+L(\eta/2-\eta_1/2)+L(1-\eta/2+\eta_1/2)
\right. \nonumber \\
\left.
-2 L(1-\eta/2)-L(1-\eta+\eta_1)+2 L(1-\eta)
\right],
\end{eqnarray}
where the occupancy probability of a sublattice is given by
\begin{equation}
\eta_1(\eta)=\frac{\eta}{2}+\sqrt{\frac{\eta^3-8\eta^2+16\eta-8}{4 \eta}}.
\end{equation}

Clearly more work, either by means of simulations or by other methods, 
is needed in order to resolve the nature of the transition. This 
notwithstanding, we do not expect good agreement except at
low or high pressures; critical points are difficult (sometimes
impossible) to capture, even approximately, with mean--field--like
theories like this one. 
Nevertheless, we think of the FM approach to be
valuable in the sense that (i) it is possible to obtain an analytic
approximation; (ii) the result obtained is compatible with one
previously reported using a different theory, and (iii) it correctly
predicts the structure of the ordered phase.

\subsection{Monocomponent and bicomponent hard cube lattice gas}
We have also applied the theory to three--dimensional systems.
In this section we present the results obtained for
the monocomponent systems of parallel hard cubes of edge lengths
$\sigma=2$ and $\sigma=6$, as well as for the binary
mixture of them. These being three--dimensional systems, 
the ordered phases that compete in the phase diagrams are 
three: smectic (one--dimensional order), columnar (two--dimensional
order) and solid (three--dimensional order). The period of
the ordering is determined from the wave--vector at which
the first divergence of the structure factor occurs. In the
first case this is $\pi$, hence the period is 2. As this is
also the period of the closest--packed phase it is the only
one that is found in stable phases. In the second case
the wave--vector is $2\pi/7$, implying a period 7; however,
in this case the closest--packed phase has period 6, so
both, period--6 and period--7 phases must be considered.
The same holds for the binary mixture.

\begin{figure}
\begin{center}
\includegraphics*[width=13cm, angle=0]{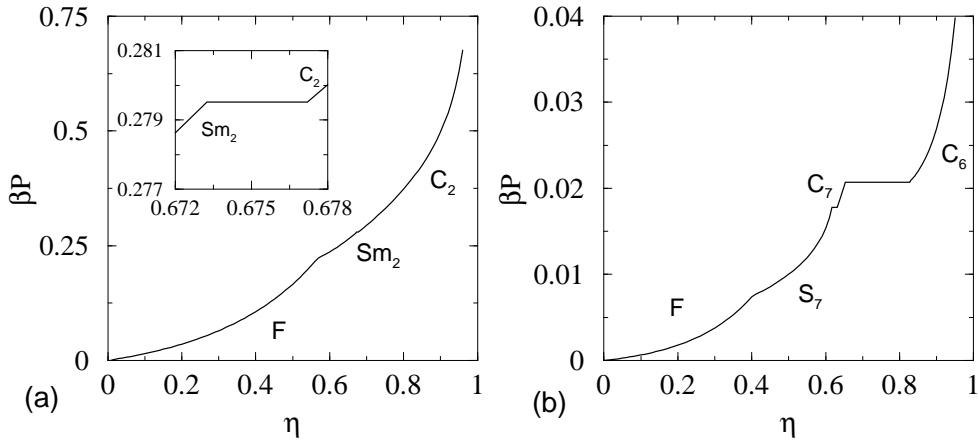}
\end{center}
\caption{\label{fig6} Equation of state obtained from FM theory
for the monocomponent hard cube lattice gas with edge lengths
$\sigma=2$ (a) and $\sigma=6$ (b). Phases are labeled F (fluid), 
Sm$_{\alpha}$ (smectic) and C$_{\alpha}$ (columnar), 
where $\alpha=2,6,7$ stands for the ordering period.
The inset in part (a) shows a very narrow first order
Sm$_2$--C$_2$ transition.
}
\end{figure}

The results for the monocomponent systems
are shown in figures \ref{fig6}a $(\sigma=2)$ and \ref{fig6}b 
$(\sigma=6)$. The first system undergoes, at $\eta_{\mathrm{c}}=0.568$,
a second order transition from a fluid phase (F) to a period--2
smectic one (Sm$_2$). At a higher
value of the packing fraction the system is found to have a
very narrow first order phase transition, where the smectic phase
at $\eta_{\mathrm{Sm}}=0.673$ coexists with a period--2 columnar 
phase (C$_2$) at $\eta_{\mathrm{C}}=0.677$ (see the inset in 
\fref{fig6}a). For the $\sigma=6$ system we find a more complex
scenario with several ordered phases, due to the proximity of
their free energies. There are six ordered phases in competition:
smectic, columnar and solid with both periods 6 and 7. Only
three of them, the two columnar and the period--7 solid, are
stable in some region. The other three are just metastable.
Again, the columnar phase is the most stable at high density
(notice the difference of the high density ordered phase of these
lattice systems with respect to that of the continuum model
(Groh and Mulder 2001)).
Upon increasing density, the sequence of transitions is as
follows: first, a second order one at $\eta_{\mathrm{c}}=0.402$ 
from a fluid phase (F) to a period--7 solid phase (S$_7$); then
a first order transition in which the S$_7$ solid coexists,
at $\eta_{S_7}=0.617$, with a period--7 columnar phase (C$_7$),
at $\eta_{\mathrm{C}_7}=0.631$; finally, another first order
transition with the C$_7$ columnar at $\eta_{\mathrm{C}_7}=0.656$
coexisting with a period--6 columnar phase (C$_6$) 
at $\eta_{\mathrm{C}_6}=0.827$. This latter phase extends up
to the closest-packing.

\begin{figure}
\begin{center}
\includegraphics*[width=12cm, angle=0]{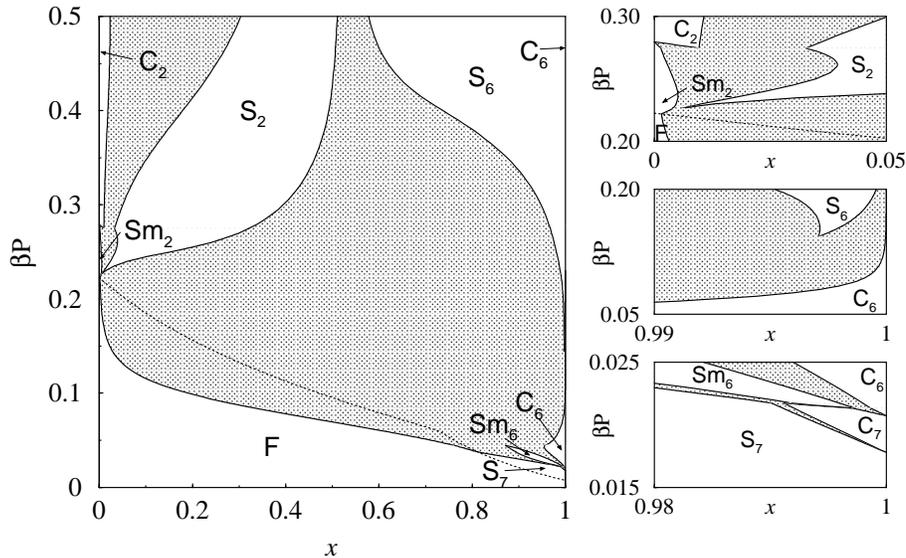}
\end{center}
\caption{\label{fig7} Phase diagram (reduced pressure, $\beta P$,
vs.\ composition, $x$) of the lattice binary mixture 
of hard cubes (size ratio 6:2). Composition is defined by
$x=\eta_{\mathrm{L}}/\eta$, where $\eta=\eta_{\mathrm{L}}+
\eta_{\mathrm{S}}$ is the total packing fraction of the large (L) 
and small (S) cubes. Phases are labeled as in \fref{fig6}.
The dotted line corresponds to the spinodal of the uniform fluid. 
For $0.81\lesssim x$ it marks a stable continuous F-S$_7$ phase
transition.}
\end{figure}

The competition of different ordered phases in the latter system
is remarkable, but it produces a real mess when the two kind
of cubes are mixed. The phase diagram for this binary mixture 
(pressure vs.\ composition) is shown in \fref{fig7}. All possible 
phases are present there. A few particularly complex regions are
shown in more detail. Remarkable features of this phase diagram
are the very wide F--C$_6$ phase separation dominating the lower
part of the diagram, and the extraordinarily narrow S$_6$--C$_6$
coexistence, extremely close to the monocomponent fluid of
large cubes. The latter tells us about the proximity of the
free energy of the solid phase to that of the columnar
in this monocomponent system (a small amount of `impurities'
breaks the translational symmetry along the columns of the
columnar phase transforming it into a solid). The former
has a very important consequence concerning the demixing
scenario for this system.

This binary mixture has been reported, from simulations
(Dijkstra and Frenkel 1994), to demix into two fluid
phases. As a matter of fact, this is, as far as we know,
the only existing evidence of an entropy--driven
fluid--fluid demixing in an additive binary
mixture. In all other systems of the same kind, e.g.\
hard spheres (Dijkstra \etal 1999) or parallel hard cubes 
in the continuum (Mart\'{\i}nez--Rat\'on and Cuesta 1998, 
1999), fluid--fluid demixing is preempted by freezing of 
(at least)
the largest component. Our results contradict this simulations
and fit in what seems to be the standard scenario of demixing
for additive binary mixtures, namely fluid--ordered phase
demixing (the fact that the ordered phase is a columnar is
just a peculiarity of the model). A more 
detailed analysis of this result can be found in
Lafuente and Cuesta (2002).

\section{Conclusions}
\label{sec5}
In this article we have proposed a formulation of FM theory
for lattice models. As it was mentioned in the introduction,
it is not possible just to carry on with Rosenfeld's original
formulation but in discrete space. This is so because of the
lack of a scaled--particle theory for lattice models. In fact,
FM theory in the continuum can be regarded as a generalization
of scaled--particle theory to inhomogeneous
phases. For this reason, we believe that the knowledge
of a lattice FM theory could guide the derivation
of a lattice scaled--particle theory. We are presently working
in that direction.

In the recent reformulations of FM theory
the scaled--particle equation of
state is no more an input but an output of the theory. Two fundamental
ideas appear as the keystone of these formulations: (i)
the exact density functional must have a consistent dimensional reduction to
any lower dimension, and (ii) the exact zero--dimensional limit
is necessary for the correct description of the freezing transition.
The latter turns out to be
a very stringent constraint for the functional and seems to contain 
most of the information needed for the construction
of a higher dimensional functional. In this context, the zero--dimensional
cavities method appears as a scheme for the construction of FM functionals.
The lattice FM theory we have proposed is the generalization of this
method suggested
by the exact form of the functional of the one--dimensional mixture
of hard rods on a lattice. Although this system has been previously studied
in the literature, 
the originality of our derivation is that (i) it is the first
time that the distinction between additive 
and nonadditive mixtures have been noticed and
resolved, and (ii) the exact functional
is written in a form in which the zero--dimensional functional arises
naturally. The cavities method can be extended to other particle shapes
and other lattice structures. This will be the content of a future work.

We have applied the functional to two and three--dimensional 
models and found very reasonable results. In two dimensions these are
comparable to those previously obtained by other means, and in three
dimensions they agree with previous simulations data, in the case
of the mixture (Lafuente and Cuesta 2002), although with
an utterly different interpretation which has yet to be confirmed.
One remarkable feature of all three--dimensional models is the very
rich and complex phase diagrams they possess (specially the mixture),
a complexity which the present theory has nevertheless allowed us
to tackle without problems.

\ack
This work is supported by project BFM2000--0004 from the
Direcci\'on General de Investigaci\'on (DGI) of the
Spanish Ministerio de Ciencia y Tecnolog\'{\i}a.

\References
\item[] Bellemans A and Nigam R K 1967 \JCP {\bf 46} 2922--34
\item[] Buschle J, Maass P and Dieterich W 2000a {\it J. Stat. Phys.}\
	{\bf 99} 273--312
\item[] \dash 2000b \JPA {\bf 33} L41--6
\item[] Cuesta J A 1996 \PRL {\bf 76} 3742--5
\item[] Cuesta J A and Mart\'{\i}nez-Rat\'on Y 1997a \PRL {\bf 78} 3681--4
\item[]\dash 1997b \JCP {\bf 107} 6379--89
\item[] Cuesta J A, Mart\'{\i}nez-Rat\'on Y and Tarazona P 2002
        \JPCM this volume
\item[] Dijkstra M and Frenkel D 1994 \PRL {\bf 72} 298--300
\item[] Dijkstra M, van Roij R and Evans R 1999 \PRL {\bf 59} 5744--71
\item[] Evans R 1979 {\it Adv. Phys.}\ {\bf 28} 143--200
\item[] \dash 1992 {\em Fundamentals of Inhomogeneous Fluids}
        ed D Henderson (Dordrecht: Kluwer) pp 85--175
\item[] Groh B and Mulder B 2001 \JCP {\bf 114} 3653--8
\item[] Hansen J P and MacDonald I R 1990 {\it Theory of Simple Liquids}
	(London: Academic Press) p 77
\item[] Lafuente L and Cuesta J A 2002 preprint
\item[] Mart\'{\i}nez-Rat\'on Y and Cuesta J A 1998 \PR E {\bf 58} R4080--3
\item[]\dash 1999 \JCP {\bf 111} 317--27
\item[] Nisbet R M and Farquhar I E 1974 {\it Physica} {\bf 73} 351--67
\item[] Percus J K 1976 {\it J. Stat. Phys.}\ {\bf 15} 505--11
\item[] \dash 1997 {\it J. Stat. Phys.}\ {\bf 89} 249--73
\item[] Robledo A 1980 \JCP {\bf 72} 1701--12
\item[] Robledo A and Varea C 1981 {\it J. Stat. Phys.}\ {\bf 26} 513--25
\item[] Rosenfeld Y 1989 \PRL {\bf 63} 980--3
\item[]\dash 1994 \PR E {\bf 50} R3318--21
\item[] Rosenfeld Y, Schmidt M, L\"owen H and Tarazona P 1996 \JPCM
        {\bf 8} L577--81
\item[]\dash 1997 \PR E {\bf 55} 4245--63
\item[] Rushbrooke G S and Scoins H I 1955 {\it Proc. Roy. Soc.}\ {\bf A230}
	74--90
\item[] Schmidt M 1999 \PR E {\bf 60} R6291--4
\item[]\dash 2000a \PRL {\bf 85} 1934--7
\item[]\dash 2000b \PR E {\bf 63} 010101(R)
\item[]\dash 2001 \PR E {\bf 63} 050201(R)
\item[] Slotte P A 1983 {\it J. Phys. C: Solid State Phys.}\ {\bf 16}
	2935--51
\item[] Soto-Campos G, Bowles R, Itkin A and Reiss H {\em J.\ Stat.\
        Phys.}\ {\bf 96} 1111--23
\item[] Tarazona P 2000 \PRL {\bf 84} 694--7
\item[] Tarazona P and Rosenfeld Y 1997 \PR E {\bf 55} R4873--6
\item[] Vanderlick T K, Davis H T and Percus J K 1989 \JCP {\bf 91} 7136--45
\item[] Widom B 1978 {\it J. Stat. Phys.}\ {\bf 19} 563--74
\item[] Zhou S 2001a \JCP {\bf 115} 2212--8
\item[]\dash 2001b \PR E {\bf 63} 061206
\endrefs
\end{document}